\documentstyle[aps,prl,preprint,floats,epsfig]{revtex}

\begin{document}

% From Pam Morehouse 2-July-2001
%
\preprint{\tighten\vbox{\hbox{\hfil CLNS 01/1744}
                        \hbox{\hfil CLEO 01-14}
}}

\title{Search for the Decay $\Upsilon$(1S)~$\to \gamma \eta^{\prime}$}  

% Your author list ***DOES NOT*** go here!
% is goes below where you are instructed to insert it...
\author{CLEO Collaboration}
% You will want to hard code the date once you are ready to submit your paper!
\date{10 July 2001}

\maketitle
\tighten

\begin{abstract} 
% Insert abstract here.

We report on a search for the radiative decay 
$\Upsilon$(1S)~$\to\gamma\eta^{\prime}$ in 
61.3 pb$^{-1}$ of data taken with the CLEO II
detector at the Cornell Electron Storage Ring.  
Three decay chains were investigated, all involving 
$\eta^{\prime}\to\pi^{+}\pi^{-}\eta$, followed by $\eta\to\gamma\gamma$,
$\eta\to\pi^{0}\pi^{0}\pi^{0}$, or $\eta\to\pi^{+}\pi^{-}\pi^{0}$.
We find no candidate events in any of the three cases and set 
a combined upper limit 
$1.6\times10^{-5}$~at 90\% C.L., significantly smaller
than the previous limit.
We compare our result to 
%prior
other 
radiative $\Upsilon$   decays,
to radiative $J/\psi$ decays, and 
to theoretical predictions.

\end{abstract}
\newpage

{
\renewcommand{\thefootnote}{\fnsymbol{footnote}}

% Insert author and address list here
% CLNS version
% /home/axp/cleoac/AuthorList/Lists/0106/JOURNALS/clns-01-1744_clns.tex
\begin{center}
S.~J.~Richichi,$^{1}$ H.~Severini,$^{1}$ P.~Skubic,$^{1}$
S.A.~Dytman,$^{2}$ V.~Savinov,$^{2}$
S.~Chen,$^{3}$ J.~W.~Hinson,$^{3}$ J.~Lee,$^{3}$
D.~H.~Miller,$^{3}$ E.~I.~Shibata,$^{3}$
% ...
\end{center}
 
\small
\begin{center}
$^{1}${University of Oklahoma, Norman, Oklahoma 73019}\\
$^{2}${University of Pittsburgh, Pittsburgh, Pennsylvania 15260}\\
$^{3}${Purdue University, West Lafayette, Indiana 47907}\\
% ...
\end{center}
\setcounter{footnote}{0}
}
\newpage

% Insert body of the text here.

The only measurement of a two-body radiative $\Upsilon$(1S) decay
is the CLEO analysis\cite{Korolkov} of 
$\Upsilon\to\gamma\pi\pi$, which was consistent with radiative
$f_{2}$(1270) production.  
In contrast, many such radiative decays
have been measured for the $J/\psi$ system\cite{PDG2000}, 
including the decay to $\gamma\eta^{\prime}$ at three times the
rate to $\gamma f_{2}$(1270).
Such radiative decays provide a ``glue-rich'' environment, which could
mean a large 
valence
gluonic component to the $\eta^{\prime}$ wave function.
Unexpectedly large rates\cite{CLEOB} 
are also observed in decays such as 
$B \to \eta^{\prime} K^{(*)}$.

In addition, there have been several theoretical predictions 
for
%KKGan - 28JUN01
%involving
the $\gamma\eta^{\prime}$ final state 
%AWarburton - 02JUL01
%involving 
that involve
non-relativistic\cite{KKKS}
or light-cone\cite{Kuhn83} approaches, with the $\eta^{\prime}$ produced
by highly virtual gluons, or models\cite{Intemann} 
involving extended vector meson dominance.  There has also been
theoretical work on the related process $\Upsilon\to\gamma\eta$\cite{Ma}
and comparisons of the $\gamma\eta$ and $\gamma\eta^{\prime}$ final
states in $J/\psi$ decay\cite{Ball}.
Further, there have been
suggestions \cite{Ball,Baier} that the decay
$J/\psi \to \gamma\eta^{(\prime)}$ might be dominated by a strong
anomaly, whereas this mechanism is suppressed for the radiative
$\Upsilon$ decays because of the more massive constituent quarks.

In this Letter we report on a search for the decay
$\Upsilon \to \gamma \eta^{\prime}$ 
%in an effort to 
and we
compare this decay mode to the $f_{2}$(1270) final state in
$\Upsilon$ decay, to the $\eta^{\prime}$ radiative decay
in $J/\psi$ decay, and to the theoretical predictions.

Our analysis used 61.3 pb$^{-1}$ of data recorded at the
$\Upsilon$(1S) resonance ($\sqrt{s} = 9.46$ GeV) with the
CLEO II detector\cite{CLEOII} 
operating at the Cornell Electron Storage Ring (CESR).
This corresponds to the production of 
$N_{\Upsilon} = (1.45 \pm 0.03)\times 10^{6}$
$\Upsilon$(1S) mesons\cite{Korolkov}.
In addition, 189 pb$^{-1}$ taken near in time to this
$\Upsilon$(1S) data but at energies just below the $\Upsilon$(4S)
were used for comparison to the four-quark continuum.
The momenta and ionization loss ($dE/dx$) 
of charged tracks were measured in a six-layer straw-tube chamber, 
a ten-layer 
precision drift chamber, and a 51-layer main drift chamber, all
operating in a 1.5~T solenoidal magnetic field. 
Photons were detected using the high-resolution electromagnetic 
calorimeter consisting of 7800 CsI crystals. The Monte Carlo simulation 
of the detector response was based upon GEANT\cite{GEANT}, and simulation
events were processed in an identical fashion to data.

Our search for 
$\Upsilon \to \gamma \eta^{\prime}$ involved the decay 
$\eta^{\prime}\to\pi^{+}\pi^{-}\eta$, followed by
$\eta\to\gamma\gamma$,
$\eta\to\pi^{0}\pi^{0}\pi^{0}$, or 
$\eta\to\pi^{+}\pi^{-}\pi^{0}$.
In order
%KKGan - 28JUN01
%Wishing 
to maximize detection efficiency and minimize possible
systematic biases, we employed a minimal number of selection criteria.
Combinatoric background is largely
suppressed by requiring reconstruction of the
three mesons: $\eta$, $\eta^{\prime}$, and $\Upsilon$.

Events were required to have the proper number of quality tracks of 
%opposite
appropriate
charges
%signs 
and at least three calorimeter 
%showers 
energy clusters
(which may or may not
be associated with the tracks), of which one had to correspond to an energy
of at least 4 GeV and be in the barrel fiducial volume
($|\cos\theta|$ $<$ 0.71).  In addition, 
%the event had to have been accepted using 
we required that the events pass
trigger requirements\cite{CLEOTrig} that were highly efficient and 
could be reliably simulated.

For reconstructing $\pi^{0}$ candidates, the photons had to have minimum
depositions of 30 (50) MeV in the 
barrel (endcap) 
regions\footnote{The endcap region 
%here 
is defined as
$0.85$$<$$ |\cos\theta| $$<$$0.95$; the region between this and the
barrel fiducial region is not used due to its poor resolution.} 
and could not be associated
with any charged track; in addition, at least one of the two photons
had to be in the barrel region. The $\gamma\gamma$ invariant mass 
had to be within
50 MeV ($\sim \pm 9 \sigma_{\pi}$) of the known $\pi^{0}$ 
mass\cite{PDG2000}; such
candidates were then kinematically constrained to that mass.
The 
%photons 
photon candidates
used in reconstructing the $\eta$ in $\gamma\gamma$
and the parent $\Upsilon$ in $\gamma\eta^{\prime}$ had
to deposit
a minimum of 60 (100) MeV in the barrel (endcap) calorimeter regions,
could not be identified as a fragment of a charged track deposition,
and had to have a lateral profile consistent with that of a photon.

Next, $\eta$ candidates were built from $\gamma\gamma$,
$\pi^{0}\pi^{0}\pi^{0}$, or $\pi^{+}\pi^{-}\pi^{0}$.  Simulation
events were used to determine the detector mass resolution for each 
of these modes: $\sigma_{\eta} = 13.4, 9.4$, and 8.2 MeV, respectively.
This was confirmed by measurements of resolution functions using
independent data samples.
Candidates had to be within $\pm 3 \sigma_{\eta}$ of the known $\eta$ mass.
In the case of the $\pi^{0}\pi^{0}\pi^{0}$ final state, no photon
could be common to more than one $\pi^{0}$ 
%building block.
combination.

Two oppositely charged tracks were then added to the $\eta$
candidate to form $\eta^{\prime}$ candidates 
that
%which 
were required
to have an invariant mass of 939$<$$m_{\pi\pi\eta}$$<$981 MeV;
this corresponds to greater than $3\sigma_{\eta^{\prime}}$ for
all three decay chains.
In the case of $\eta\to\gamma\gamma$, a charged track was
rejected if its momentum, $p$, from the drift chamber matched
its energy, $E$, as measured in the calorimeter as
0.85 $<$$E/p$$<$~1.05; this further suppressed QED backgrounds in this
mode.

Finally, $\Upsilon$ candidates were formed by adding the high energy
photon ($E$$>$4 GeV) to the $\eta^{\prime}$ candidate, being sure that this photon
was not already used in reconstructing the event.  To be considered,
such a candidate had to have an invariant mass within $\pm 300$ MeV
of $\sqrt{s} = m_{\Upsilon}$, which is roughly three times the detector
resolution as obtained from our simulations.
Although,  in general, multiple candidates per event were not restricted,
there were two exceptions:
(i) in the case of $\eta\to\pi^{0}\pi^{0}\pi^{0}$,
if two $\eta$ candidates shared more than four photons, the candidate
with the better combined $\chi^{2}$ for mass fits to 
the three $\pi^{0}$ candidates was accepted; and (ii) in the 
case of $\eta\to\pi^{+}\pi^{-}\pi^{0}$, if two candidates for the neutral
pion shared a daughter photon, the one with the better fit to the
$\pi^{0}$ mass was taken.

After these 
highly-efficient
procedures were applied, we found {\it no} candidates in either
the $\Upsilon$ or continuum data samples.

From Monte Carlo simulations, the overall
efficiencies, $\epsilon_{i}$, were determined to
be $(31.8 \pm 1.8)\%, (15.0 \pm 1.6)\%$, and $(21.1 \pm 1.4)\%$
for the decay chains ending in 
$\eta\to\gamma\gamma$,
$\eta\to\pi^{0}\pi^{0}\pi^{0}$, and 
$\eta\to\pi^{+}\pi^{-}\pi^{0}$, respectively.  The
uncertainties here include the statistics of the Monte
Carlo samples and our estimates on possible systematic biases,
which we discuss below.  Including the branching fractions
for the $\eta^{\prime}$ and $\eta$ decays\cite{PDG2000}
and their uncertainties gave
${\cal B}(\eta^{\prime}\to\eta\pi^{+}\pi^{-})\cdot
\sum [\epsilon_{i}{\cal B}_{\eta,i}] = (9.7 \pm 0.5)\%$.

The major
%KKGan - 28JUN01
%Non-negligible 
sources of possible systematic bias 
in our efficiency calculation from
modeling are shown in Table \ref{systerr}.
The uniformity and definition of the fiducial 
volume of the barrel calorimeter ($\pm$2.2\%) 
%was checked using
%various known processes such as Bhabha scattering.
relates to our correctly modelling the detector response to
the proper angular distribution for radiative $\Upsilon$ decay.
Uncertainties in charged track reconstruction ($\pm$0.5\% per track),
reconstruction of $\pi^{0}$ and $\eta$ mesons from photons\cite{ProcarioBalest}
($\pm$3\% per meson), and trigger effects ($\pm$2.5\%) 
were determined from previous
detailed CLEO studies 
of low multiplicity $\tau$-pair and $\gamma\gamma$ events.  
Our ability to model the $E/p$ requirement in the $\gamma\gamma$ final
state was assessed using charged pions from $K_{S}$ decays and 
assigned a 2.1\% uncertainty.
Detector stability was monitored by comparing the reconstruction
efficiencies for the $\eta$, $\eta^{\prime}$ and $\Upsilon$ as a function
of time; only in the final state $\pi^{+}\pi^{-}\pi^{0}$ was any
variation noted, for which we have assigned a 
3\%
%KKGan - 28JUN01
%$\pm$3\% 
uncertainty.
Shower leakage and other calorimeter effects make 
the mass distribution for $\Upsilon$ candidates asymmetric; based on
CLEO experience
with exclusive radiative $B$ meson decays\cite{Savinov} we have
assigned a 2\% uncertainty regarding our ability to model these effects.
These uncertainties were added in quadrature,
along with the statistical uncertainty associated with the size of
Monte Carlo samples, to obtain
the overall systematic uncertainty in the efficiencies. 

\begin{table}[hbt]
\caption{Systematic uncertainty contributions, as relative percentages,
to the 
efficiency for the studied decay modes. 
%These were added in quadrature.
The combined uncertainties were obtained using quadrature addition.
}
\begin{center}
\begin{tabular}{|l|c|c|c|}
%\hline
Uncertainty source	& $\gamma\gamma$ & $\pi^{0}\pi^{0}\pi^{0}$ & $\pi^{0}\pi^{+}\pi^{-}$  	\\
\hline
Fiducial requirements	
& 2.2		& 2.2		& 2.2		\\
Track reconstruction	        
& 1		& 1		& 2		\\
$\eta,\pi^{0}$ reconstruction from $\gamma\gamma$	
& 3		& 9		& 3		\\
Trigger	simulation
& 2.5		& 2.5		& 2.5		\\
$E/p$ criterion
& 2.1		& ~-~		& ~-~		\\
Reconstruction stability 
&~-~            &~-~            & 3    		\\
$\Upsilon$ mass distribution 
& 2		& 2		& 2		\\
\hline
Monte Carlo statistics  
& 1.9		& 3.5		& 2.5		\\
\hline
\hline
Combined uncertainty		
& 5.8		& 10.4		& 6.7		\\
%\hline
\end{tabular}
\end{center}
\label{systerr}
\end{table}

Given that we found zero candidates we applied 
%the 
a
%standard 
frequentist
%(as opposed to Bayesian) 
approach\cite{Cousins}, 

\begin{equation}
N_{true} = N_{\Upsilon} \cdot 
{\cal B}(\Upsilon\to\gamma\eta^{\prime})\cdot
{\cal B}(\eta^{\prime}\to\eta\pi^{+}\pi^{-})\cdot
\sum[\epsilon_{i}{\cal B}_{\eta,i}]~~,
\end{equation}

%Reordered as suggetsed by KKGan - 28JUN01
implying that the
mean actual number of $\gamma\eta^{\prime}$ 
events, $N_{true}$, is less than 2.3 
at 90\% C.L.  
To include systematic effects,
we performed a large number of ``toy'' Monte Carlo experiments in 
which we used values of $N_{true}$ distributed in accordance with
Poisson statistics and used values of efficiencies, branching fractions, and
$N_{\Upsilon}$ distributed as Gaussian functions with their associated
uncertainties.  From the resulting distributions we found
90\% C.L. limits for 
${\cal B}(\Upsilon\to\gamma\eta^{\prime})$ of 
$2.9 \times 10^{-5}, 
7.6 \times 10^{-5}$, and 
$7.5 \times 10^{-5}$
for the final $\eta$ states of 
$\gamma\gamma$,
$\pi^{0}\pi^{0}\pi^{0}$ and
$\pi^{0}\pi^{+}\pi^{-}$, respectively. For the sum of the
three modes we found

\begin{equation}
{\cal B}(\Upsilon\to\gamma\eta^{\prime}) < 1.6 \times 10^{-5},
\end{equation}

again at 90\% C.L.
%If we exclude incorporation of the systematic uncertainties the
Without systematic uncertainties the
limit decreases by less than 1\% of its value.
%This 
Our result
can be compared to the previous Crystal Ball limit\cite{xbal}
of $1.3\times 10^{-3}$.

\begin{figure}[htb]
\centerline{\epsfig{file=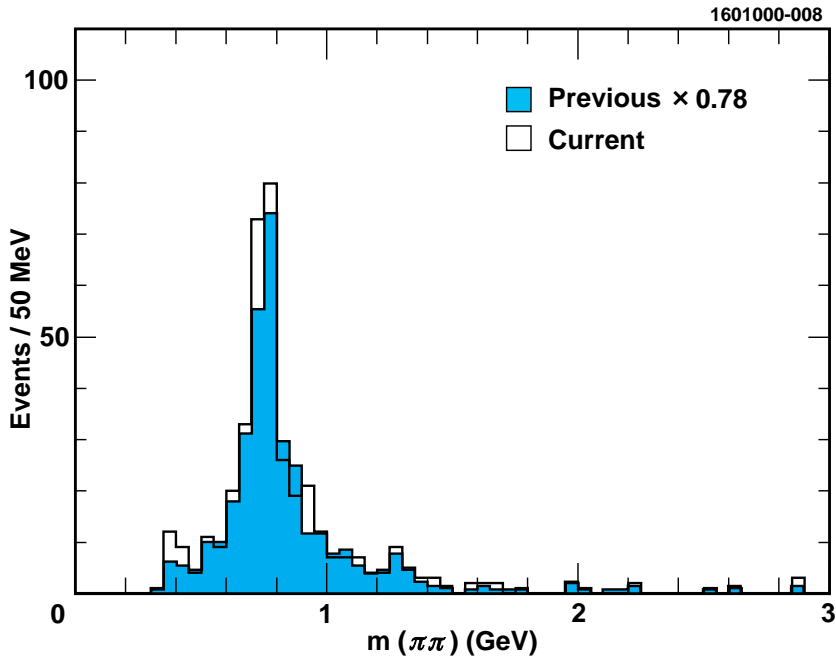,
%     height=4.0in}}
    height=4.0in,
    bbllx=-50, bblly=300,bburx=400,bbury=700}}
\vspace{10pt}
\caption{Mass spectrum for $\gamma\pi^{+}\pi^{-}$ events for data
collected at the $\Upsilon$(1S) resonance from this analysis and the
prior CLEO analysis, which has been scaled by 0.78 so that
both represent the same integrated luminosity.}
\label{IlyaGeorg}
\end{figure}

\begin{figure}[htb]

\centerline{\epsfig{file=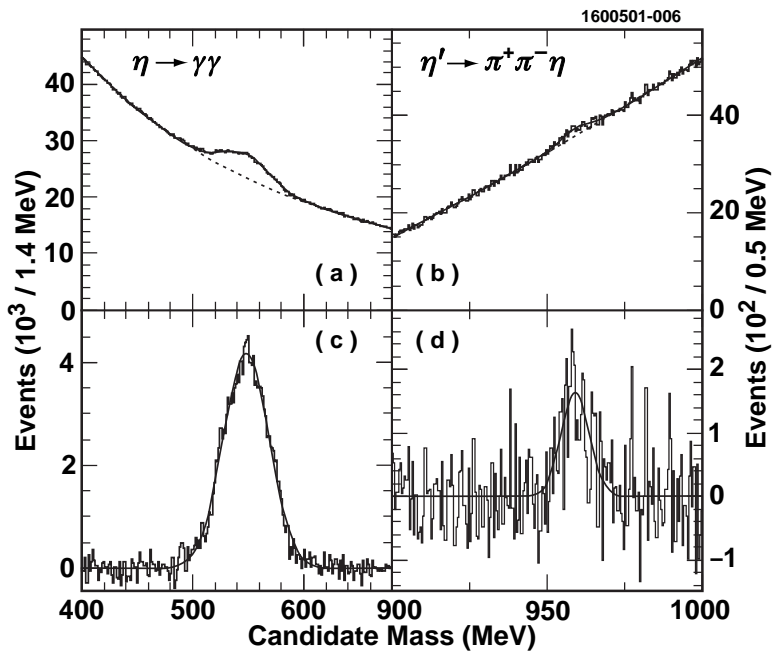,
%     height=4.0in}}
    height=4.0in,
    bbllx=80, bblly=300,bburx=550,bbury=750}}
%\vspace{10pt}
\caption{The $\eta\to\gamma\gamma$ and 
$\eta^{\prime}\to\pi^{+}\pi^{-}\eta$ invariant mass distributions from
data taken at or near the $\Upsilon$(4S). The upper plots (a and b) give the 
invariant mass distributions (histograms), which are each fit with 
the sum (solid lines) of a polynomial 
background (dashed lines) and a Gaussian signal. The
lower plots (c and d) show the distributions after subtraction of the polynomial 
background, along with the Gaussian fits.  The scale on the right is for
plots b and d.}

\label{valeragg}

\end{figure}

To show that we could use our data to observe known final states
that have high energy photons and a small number of charged tracks,
we first reproduced the $\gamma\pi^{+}\pi^{-}$ spectrum 
previously reported by 
CLEO\cite{Korolkov}.  We observed the same features 
%($\gamma\rho$ and
%$\gamma\phi$ from the continuum; an enhancement at 1270 MeV) 
at 
%the same 
similar
magnitudes as that study,
as demonstrated in Figure~\ref{IlyaGeorg}. 
We also applied our same selection criteria, with the exception of
requiring a high energy photon, to data samples taken at the
$\Upsilon$ and at or near the $\Upsilon$(4S) and
found $\pi^{0}, \eta$, and $\eta^{\prime}$ candidates at the expected 
rates\cite{PDG2000}; an example of this is shown in Figure~\ref{valeragg}. 

To compare our result to other radiative decays we use the established
$J/\psi$ branching fractions\cite{PDG2000} and the prior CLEO 
work\cite{Korolkov} for $\Upsilon\to\gamma\pi^{+}\pi^{-}$.  For the
latter,  we assume the enhancement at 1270 MeV is all attributable to 
$f_{2}$ production and that ${\cal B}(f_{2}\to\pi^{+}\pi^{-})$ is
2/3 of 84.7\% to obtain 
${\cal B}(\Upsilon\to\gamma f_2) = (8.2 \pm 3.6) \times 10^{-5}$.
We then form the ratio $R(V) = {\cal B}(V\to\gamma\eta^{\prime})/
{\cal B}(V\to\gamma f_{2})$ and calculate $R(J/\psi) = 3.1 \pm 0.4$ whereas
we obtain a 
90\% C.L. limit of $R(\Upsilon) < 0.26$.  Here we have made no attempt to
consider possible correlations between the measurements forming the ratios.
Clearly the situation is 
different for 
%these two cases.
$J/\psi$ and $\Upsilon$(1S).

The models of K\"orner, K\"uhn, Krammer, and Schneider\cite{KKKS,Kuhn83}
use highly virtual gluons to form the final state
mesons; 
%they 
these models
predict
${\cal B}(\Upsilon\to\gamma\eta^{\prime}) = 20 \times 10^{-5}$,
but are sensitive to the running of $\alpha_{s}$ between the charm and
bottom mass scales.  The recent compilations\cite{PDG2000} of $\alpha_{s}$ 
would tend to
lower this prediction\cite{KuhnPriv} to $5-10 \times 10^{-5}$, 
still significantly
larger than our new limit.  
The Intemann model{\cite{Intemann}, using extended 
vector meson dominance, gives bounds of
$5.3 \times 10^{-7} \leq {\cal B}(\Upsilon\to\gamma\eta^{\prime})
\leq 2.5 \times 10^{-6}$,
with the limits corresponding to  
%KKGan - 28JUN01
%\leq 2.5 \times 10^{-6}$.  
%The two limits come from not knowing whether 
the amplitudes
from the virtual vector mesons 
interfering
%KKGan - @*JUN01
%interfere 
destructively or constructively.
Although
one of the inputs to the theory is outdated, it is clear that our experiment does not
have the sensitivity to test this prediction.
Using NRQCD matrix elements for the $\Upsilon$ and twist-2 and twist-3
amplitudes for the gluons, Ma\cite{Ma} obtains the related
branching fraction 
${\cal B}(\Upsilon\to\gamma\eta) \approx 1.2\times 10^{-7}$, again
below our present sensitivity.

A more robust prediction of the model of K\"orner {\it et al.} is for the double ratio
of rates,\footnote{This ratio is constructed 
from Table IV  of Ref.\cite{KKKS}} 
which is independent of $\alpha_{s}$:

$${\cal R}= \frac
{{\cal B}(\Upsilon\to\gamma\eta^{\prime})}
{{\cal B}(\Upsilon\to\gamma f_{2})}
\times
\frac
{{\cal B}(J/\psi \to \gamma f_{2})}
{{\cal B}(J/\psi \to \gamma\eta^{\prime})}
= \frac{0.11}{0.24} = 0.46.$$

Using our result, we obtain 
an upper limit of 
0.09 for this double ratio at 90\% C.L.; the
probability that our result 
%could fluctuate to 
is consistent with
0.46 is 0.6\%.

In summary, we have searched for the decay $\Upsilon\to\gamma\eta^{\prime}$
with the decay mode $\eta^{\prime}\to\eta\pi^{+}\pi^{-}$ and three decay
modes of the $\eta$.  
Using simple, loose selection criteria,
%We 
we found no candidates and set the 90\% C.L. limit
of ${\cal B}(\Upsilon\to\gamma\eta^{\prime}) < 1.6\times10^{-5}$.  
This is significantly
small when compared to other radiative decays of heavy vector
mesons and smaller than 
%would be predicted from 
%theories 
theoretical predictions
that use highly virtual gluons in
forming final state mesons. 
%would predict.

% January 2001 PRL acknowledgments

We gratefully acknowledge the effort of the CESR staff in providing us with
excellent luminosity and running conditions.
This work was supported by 
the National Science Foundation,
the U.S. Department of Energy,
the Research Corporation,
the Natural Sciences and Engineering Research Council of Canada
and the Texas Advanced Research Program.

\end{document}